\documentclass[twocolumn,preprintnumbers,amsmath,amssymb,prb]{revtex4}
\usepackage{graphicx}
\begin{document}
\draft
\title{Proximity effects induced by a gold layer on La$_{0.67}$Sr$_{0.33}$MnO$_3$ thin films}
 \normalsize
\author{R. Bertacco, S. Brivio, M. Cantoni, A. Cattoni, D. Petti, M. Finazzi, and F. Ciccacci}
\address{LNESS - CNISM - Dipartimento di Fisica, Politecnico di Milano, Via Anzani 42, 22100 Como, Italy}
\author{A. A. Sidorenko, M. Ghidini, G. Allodi, and R. De Renzi}
\address{Dipartimento di Fisica, Universit\`{a} di Parma,  Viale delle Scienze, 43100 Parma, Italy}
\date{\today}

\bigskip
\begin{abstract}
\noindent We report about La$_{0.67}$Sr$_{0.33}$MnO$_3$ single
crystal manganite thin films in interaction with a gold capping
layer. With respect to uncoated manganite layers of the same
thickness, Au-capped 4~nm-thick manganite films reveal a dramatic
reduction ($\simeq 185$~K) of the Curie temperature $T_\mathrm{C}$
and a lower saturation low-temperature magnetization $M_0$. A
sizeable $T_\mathrm{C}$ reduction ($\simeq 60$~K) is observed even
when an inert SrTiO$_3$ layer is inserted between the gold film and
the 4~nm-thick manganite layer, suggesting that this effect might
have an electrostatic origin.
\end{abstract}

\pacs{75.47.Lx, 75.70.Cn}


\maketitle

\newpage

The perovskite manganites of composition La$_{1-x}$A$_x$MnO$_3$ (A =
Ca, Sr, Ba) have attracted a considerable interest because of the
interplay between spin, charge, orbital, and lattice degrees of
freedom in these materials, leading to a large variety of magnetic
and electronic properties, such as colossal
magnetoresistance.\cite{Coey} The half-metallic character of
manganites with A = Sr, Ca is particularly relevant for applications
in spin electronics where sources and detectors of highly polarized
electron currents are required.

The electric and magnetic properties of manganites are extremely
sensitive to the concentration of free carriers. These properties
can be varied not only by hole-doping the material, but also by
designing a heterostructure where light or an externally applied
electric field might effectively modulate the carrier
concentration.\cite{Ahn} Attempts to incorporate \textit{p}-doped
manganites in heterostructures with \textit{n}-type conductors have
demonstrated the possibility to control the carrier concentration
and tune the intrinsic magnetic properties of the manganite layer
through the modulation of interfacial electronic bands either with
external bias voltages or carrier
injection.\cite{Katsu,Tanaka,Tiwari,Zhou} Modulation of the
transport and magnetic properties of epitaxial manganite films has
also been obtained in field-effect devices, where a voltage is
applied to a metal\cite{Ogale,Xie,Hu} or
ferroelectric\cite{Grishin,Mathews,W.Wu,T.Wu,Hong,Kanki,Eerenstein}
gate electrode.

The presence of an interface represents, by itself, a significant
perturbation of the electronic properties of perovskite
manganites.\cite{Ahn} The interface with a metal is known to affect
the chemical and electronic environment of strongly correlated
oxides (a wide group of materials where electron-electron
correlations play a fundamental role and that includes manganites),
and can change basic electronic parameters, such as the Hubbard
energy $U$, the electronic bandwidths, or the exchange
energies.\cite{Altieri} Nevertheless, the impact of the creation of
a manganite/metal interface has been scarcely
investigated,\cite{Mieville} although its understanding is a
fundamental step in the study of the effects taking place when other
perturbations, such as electric fields, are applied to the
heterostructure.

In the following, we report about the dramatic effects that
depositing a metal layer on a La$_{0.67}$Sr$_{0.33}$MnO$_3$ (LSMO)
thin film has on the manganite magnetic properties. Since  bulk LSMO
is metallic at the $x=0.33$ Sr doping concentration, band bending
cannot take place throughout an extensive space charge region,
contrary to what happens in the insulating manganite phases.
Therefore, the interface with a different material is expected to
influence the LSMO electronic and magnetic properties only over a
very short length scale. It is thus essential to grow high quality
thin LSMO layers forming very sharp interfaces with the metal. A
noble metal is the most obvious choice as a partnering material, in
the attempt to reduce chemical interdiffusion and interface
reactions. Aluminium, for instance, drains oxygen from the
manganite, altering its stoichiometry and strongly affecting the
junction resistance vs. temperature dependence.\cite{Plecenik}

Thin single crystal LSMO films (thickness $t_{\mathrm{LSMO}}$
between 4 and 12~nm) were deposited on SrTiO$_3$ (STO) by pulsed
laser deposition using a tripled Nd:YAG laser (wavelength = 355~nm).
The laser fluence was 1.4~J/cm$^2$ with a repetition rate equal to
2~Hz. The sample temperature and oxygen pressure were 973~K and
0.29~mbar, respectively. The surface roughness was characterized by
Atomic Force Microscopy (AFM) and proved to be $\simeq 0.2$~nm
root-mean-square, with a very small number of droplets covering less
than 1\% of the sample area. The LSMO deposition rate was calibrated
by X-ray reflectometry, so that the estimated error in the thickness
is lower than 0.2~nm.

\begin{figure}
\includegraphics[width=0.38\textwidth]{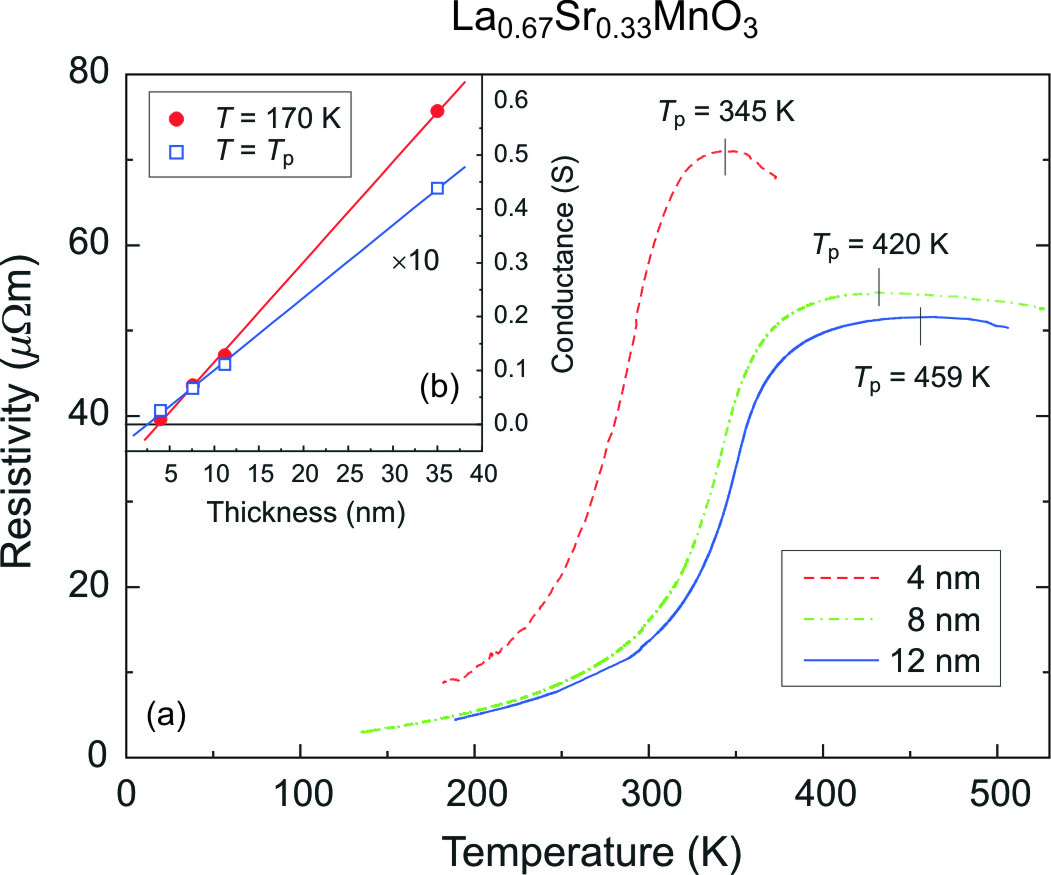}
\caption{\label{fig:one}(color online) (a) Temperature dependence of
the resistivity in thin LSMO films with thickness equal to 4, 8, and
12~nm, measured in the four-probe configuration. (b) LSMO
conductance as a function of the LSMO film thickness, at $T=170$~K
and at $T=T_\textrm{p}$. The lines in the inset are linear fits to
the data points.}
\end{figure}

All the LSMO thin films reproduce the characteristic
metal-to-insulator transition [see Fig.~\ref{fig:one}(a)] typical of
bulk LSMO. The temperature $T_\mathrm{p}$ at which the resistivity
maximum is observed decreases with the LSMO film thickness, as
discussed in Ref.~\onlinecite{Hong}. From the thickness dependence
of the LSMO layer conductance [Fig.~\ref{fig:one}(b)] one can see
that LSMO becomes electrically insulating below a critical thickness
of the order of $\approx 3$~nm, in agreement with
Ref.~\onlinecite{Hong}. Anyway, the presence of a maximum in the
resistivity vs. temperature dependence in Fig.~\ref{fig:one}(a)
ensures that even the LSMO thin films with $t_{\mathrm{LSMO}}=4$~nm
cannot be considered as electrically dead.\cite{Hong}

\begin{figure}
\includegraphics[width=0.33\textwidth]{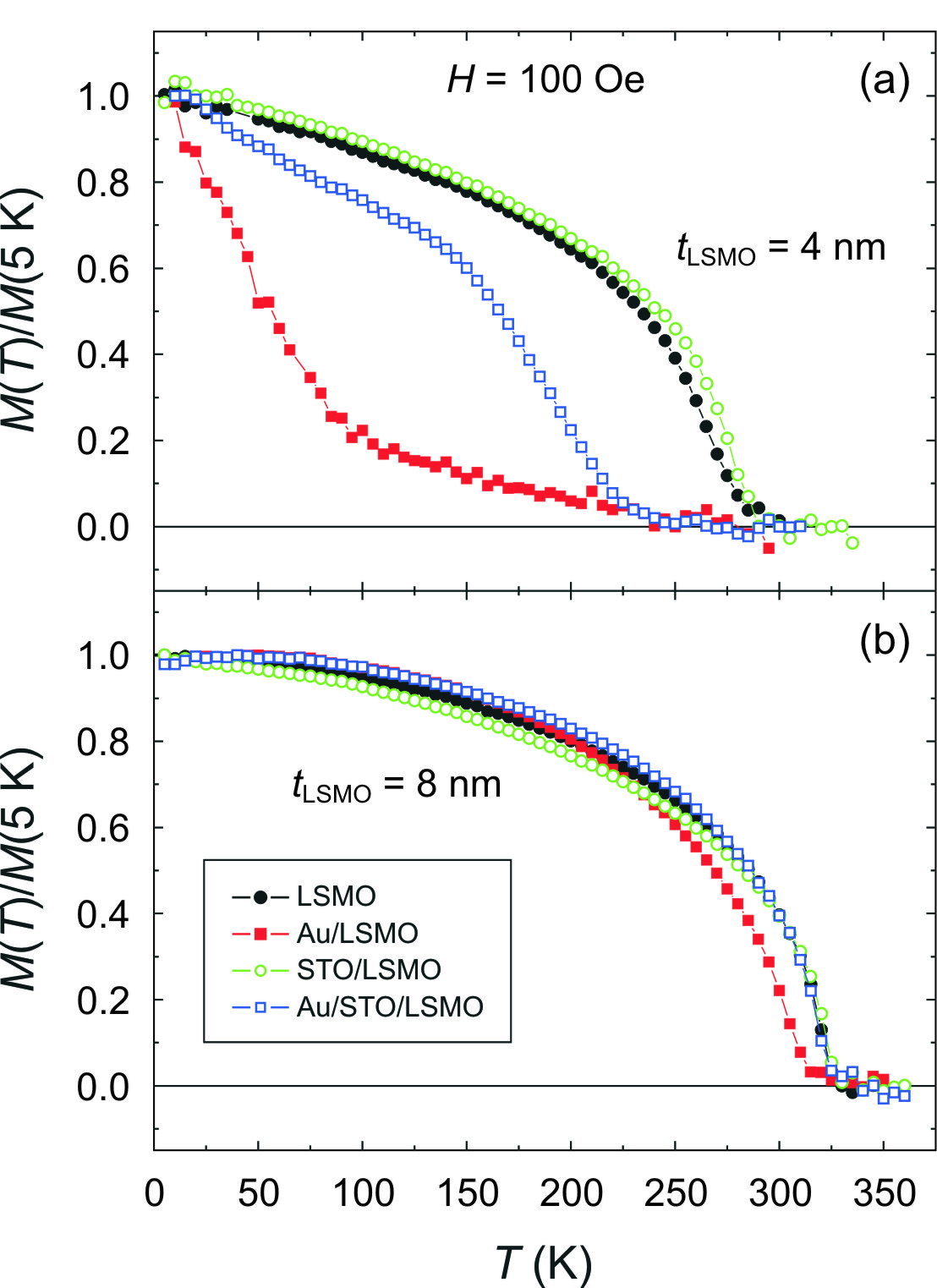}
\caption{\label{fig:two}(color online) Normalized magnetization vs.
temperature curves measured by SQUID for different heterostructures:
(a) LSMO thickness $t_{\mathrm{LSMO}}=4$~nm or (b)
$t_{\mathrm{LSMO}}=8$~nm. The STO and Au thickness is 2~nm.}
\end{figure}

The sample magnetization $M$ as a function of the temperature $T$
has been determined by a Superconducting Quantum-Interference Device
(SQUID) and is reported in Fig.~\ref{fig:two} for LSMO thin films
with thickness $t_{\mathrm{LSMO}}=4$~nm [Fig.~\ref{fig:two}(a)] and
$t_{\mathrm{LSMO}}=8$~nm [Fig.~\ref{fig:two}(b)]. The Curie
temperature $T_\mathrm{C}$ is obtained as the temperature
corresponding to the higher temperature inflection point of $M(T)$.
In agreement with the observed trend of $T_\mathrm{p}$,
$T_\mathrm{C}$ is lower than the bulk value and decreases with the
film thickness. We also note that the $T_\mathrm{p}$ reduction we
observe by varying $t_{\mathrm{LSMO}}$ from 8 to 4~nm is more
pronounced than the reduction of $T_\mathrm{C}$. This indicates that
the magnetic and electric transitions do not coincide, as
highlighted in Ref.~\onlinecite{Bertacco2005}, and suggests that in
thin LSMO films the transport properties are more affected than the
magnetic ones.

Subsequently, we have studied the effect of depositing a thin
(2~nm-thick) gold layer on top of LSMO films prepared as described
above. Au has been evaporated from a Knudsen cell (pressure during
deposition $\simeq 10^{-9}$~mbar). AFM does not measure any
degradation of the surface roughness after Au deposition. Transport
measurements across the Au/LSMO interfaces evidence an ohmic (non
rectifying) behavior.

The effects of the Au capping on both the LSMO Curie temperature and
the saturation magnetization $M_0$ at low temperature ($T=5$~K) are
noteworthy, as seen from Fig.~\ref{fig:two} and
Table~\ref{tab:table1}, where values of $T_\mathrm{C}$, $M_0$, and
$\Delta T_\mathrm{C}$ (the shift of the Au-capped LSMO layer Curie
temperature with respect to the same thickness uncapped layer) are
listed.  A sizeable reduction of $T_\mathrm{C}$ and $M_0$ is
observed for $t_{\mathrm{LSMO}}=4$~nm, while for
$t_{\mathrm{LSMO}}=8$~nm the $T_\mathrm{C}$ and $M_0$ reduction is
much smaller but still evident (see Fig.~\ref{fig:two}). No
significant modifications are evidenced between uncapped and
Au-capped LSMO thin films with $t_{\mathrm{LSMO}}=12$~nm.

We have performed a X-ray Photoemission Spectroscopy (XPS) study of
Au-capped LSMO samples as a function of the gold layer thickness, to
check for interdiffusion of chemical species at the interface by
monitoring the core photoemission lines and/or Auger
peaks.\cite{Bertacco.unpublished} The shape and position of the
characteristic peaks from LSMO are essentially unchanged upon Au
deposition, indicating the absence of relevant chemical reactions at
the interface, at variance with what observed for Al on
LSMO.\cite{Plecenik} Our results are consistent with recent x-ray
absorption and circular dichroism spectroscopy data demonstrating
that the deposition of Au on LSMO has no effect on the shape of the
spectra collected at the Mn $L_{2,3}$ edges.\cite{Stadler} No
sizable interdiffusion of chemical species was detected beyond the
instrumental sensitivity (about 10\%) from the XPS peaks intensity
trend vs. Au coverage. However, it must be noticed that the Curie
temperature in LSMO is extremely sensitive to even small variations
(a few percent) of the oxygen concentration,\cite{Guevara} so
chemical effects on $T_\mathrm{C}$ or $M_0$ cannot be completely
ruled out by our XPS analysis.

\begin{table}
\caption{\label{tab:table1} Curie temperature $T_\mathrm{C}$, Curie
temperature shift $\Delta T_\mathrm{C}$ and low temperature
magnetization $M_0$ in LSMO thin films with and without Au and/or
STO capping.}
\begin{ruledtabular}
\begin{tabular}{rcccc}

Sample&$t_{\mathrm{LSMO}}$&$T_\mathrm{C}$&$\Delta T_\mathrm{C}$&$M_0$\\
&(nm)&(K)&(K)&(emu/cm$^3$)\\

\hline

LSMO&$4\pm0.1$&$280\pm 5$&&$420\pm 70$\\
Au/LSMO&$4\pm0.1$&$95\pm 10$&$185\pm 11$&$150\pm 100$\\
STO/LSMO&$4\pm0.1$&$285\pm 5$&$-5\pm 7$&$500\pm 70$\\
Au/STO/LSMO&$4\pm0.1$&$220\pm 10$&$60\pm 11$&$420\pm 70$\\

LSMO&$8\pm0.1$&$325\pm 2$&&$590\pm 65$\\
Au/LSMO&$8\pm0.1$&$315\pm 2$&$10\pm 3$&$420\pm 65$\\
STO/LSMO&$8\pm0.1$&$325\pm 2$&$0\pm 3$&$480\pm 65$\\
Au/STO/LSMO&$8\pm0.1$&$325\pm 2$&$0\pm 3$&$520\pm 65$\\

LSMO&$12\pm0.1$&$345\pm 2$&&$490\pm 50$\\
Au/LSMO&$12\pm0.1$&$340\pm 2$&$5\pm 3$&$450\pm 50$\\

\end{tabular}
\end{ruledtabular}
\end{table}

In order to suppress interdiffusion and interface chemical
reactions, we have inserted a thin (2~nm-thick) STO film between the
Au capping and the LSMO layer. STO is an insulating oxide with a
very low lattice mismatch (0.8\%) with respect to LSMO. Therefore,
possible strain effects on the LSMO layer are minimized. Strain, in
fact, is known to significantly alter the electronic properties of
perovskite manganites.\cite{Millis,Ahn.Nature,Thiele,Tebano} As one
can see from Fig.~\ref{fig:two} and Table~\ref{tab:table1}, the
effects induced on both $T_\mathrm{C}$ and $M_0$ by a STO overlayer
without gold capping are negligible. Therefore, the STO overlayer
can be considered as inert.

The magnetic characterization performed by SQUID reveals that the
Au/STO/LSMO multilayer with $t_{\mathrm{LSMO}}=8$~nm and the
corresponding uncapped LSMO film with the same thickness have,
within the experimental uncertainty, the same $T_\mathrm{C}$ and
$M_0$ values (see Table~\ref{tab:table1}). Conversely, the
Au/STO/LSMO heterostructure with $t_{\mathrm{LSMO}}=4$~nm
surprisingly still presents a large reduction ($\simeq60$~K) of
$T_\mathrm{C}$.

As mentioned above, the manganite $T_\mathrm{C}$ reduction induced
by the Au capping through the STO spacer at $t_{\mathrm{LSMO}}=4$~nm
is not likely to be due to interdiffusion or strain. This conclusion
is corroborated by the fact that, in this sample, $M_0$ is unaltered
with respect to the value measured on uncapped LSMO layers. However,
the presence of the STO spacer does not exclude space charge effects
very close to the interface, although it may attenuate them. Local
charge variations directly affect the electronic properties of LSMO.
Apparently, magnetism is an extremely sensitive probe of these
localized variations, whereas the usual voltage-drop signature of a
conventional Schottky barrier is strongly suppressed in this case by
the reduced spatial extent, as demonstrated by the measured ohmic
behavior of the Au/LSMO junctions. Interfacial modifications are
naturally most effective on very thin films and this might be the
reason why the effects of the Au capping layer are reduced in the
samples with $t_{\mathrm{LSMO}}=8$~nm.

We would like to underline that the $T_\mathrm{C}$ reduction in Au-
and Au/STO-capped LSMO thin films is very high compared to the
field-induced $T_\mathrm{C}$ shifts (a few K) typically observed in
field effect devices,\cite{Kanki} and of the same order of magnitude
of the $T_{\mathrm{p}}$ shifts (up to 130~K) induced by charge
injection in \textit{p-n} structures.\cite{Tiwari} This suggests
that thin manganite films might be particularly suitable for devices
where transport and magnetic properties could be tuned by the
externally applied electric fields.

In summary, we have observed that capping
La$_{0.67}$Sr$_{0.33}$MnO$_3$ thin films with gold has sizeable
effects on the manganite Curie temperature and saturation
magnetization at low temperature. These effects consist in a strong
reduction films of $T_\mathrm{C}$ and $M_0$ in Au-capped LSMO thin
films with respect to uncoated LSMO films of the same thickness. For
the thinnest (4~nm-thick) investigated LSMO film, a smaller but
significant $T_\mathrm{C}$ reduction ($\simeq60$~K) is observed even
when an inert SrTiO$_3$ spacer is inserted between the gold film and
the manganite layer. This finding suggests that the proximity
effects induced by gold on the LSMO layer magnetic properties might
have, at least partially, an electrostatic origin.
\\
\\
\indent This work has been funded by Consorzio Nazionale
Interuniversitario per le Scienze Fisiche della Materia (CNISM) as
part of a Progetto d'Innesco della Ricerca Esplorativa 2005, and
partly by Netlab NanoFaber and FPVI STREP OFSPIN.

\end{document}